\documentclass{svproc}
\usepackage{url}

\usepackage{graphicx} 

\begin{document}
\mainmatter              
\title{Input from the charm threshold for the measurement of $\gamma$}
\titlerunning{Charm threshold input}  
%
\author{P. K. Resmi\\ (Acknowledge the erstwhile CLEO collaboration members for the privilege of using the data for the results presented)}
\authorrunning{P. K. Resmi et al.} 
%
\tocauthor{P. K. Resmi, on behalf of CLEO collaboration}
\institute{Indian Institute of Technology Madras, Chennai, India\\
\email{resmipk@physics.iitm.ac.in}}

\maketitle              

\begin{abstract}
A brief overview of the inputs from charm threshold that are essential to the determination of the Unitarity Triangle angle $\gamma$ is presented. The focus is on the measurements of four-body final states that have not previously been considered: $D^0\to K_{\rm S}^0 \pi^+ \pi^- \pi^0 $ and $D^0 \to \pi^+ \pi^- \pi^+ \pi^-$.
\end{abstract}
\section{Introduction}

Among the three CKM \cite{C,KM} angles, the uncertainty on $\gamma$ is much worse than that on $\beta$. This is due to the small branching fraction of decays sensitive to $\gamma$. An improved measurement of $\gamma$ is essential for testing the standard model description of $\mathit{CP}$  violation. The decays $B^{\pm}\to DK^{\pm}$, where $D$ indicates a neutral charm meson reconstructed in a final state common to both $D^{0}$ and $\bar{D^{0}}$, provide $\mathit{CP}$ -violating observables and they can be used for measuring $\gamma$ by analysing data collected at detectors such as BaBar, Belle, LHCb or the future Belle II experiment. These are tree-level decays and hence the theoretical uncertainty is $\mathcal{O}(10^{-7})$ \cite{Brod}. 

There are different methods of measuring $\gamma$ depending on the $D$ final state. If both $D^0$ and $\bar{D^0}$ decay to a $\mathit{CP}$  eigenstate such as $K_{\rm S}^0\pi^0$ or $K^+K^-$, then the GLW formalism \cite{GLW1,GLW2} is used for the measurement. When the $D$ meson decays into Cabibbo favoured and doubly Cabibbo suppressed final states like $K^{\mp}\pi^{\pm}$, the ADS method \cite{ADS} is used for the extraction. In these methods, asymmetry parameters and charge-averaged rates are measured from which $\gamma$ is extracted. Multibody $D$ decays like $K^{\mp}\pi^{\pm}\pi^0$ or $K^{\mp}\pi^{\pm}\pi^{\pm}\pi^{\mp}$ can be analysed using this method if coherence factor $\kappa$ is known~\cite{k3pi}. The GGSZ framework \cite{GGSZ} is used when $D$ decays to multibody self-conjugate final states like $K_{\rm S}^0\pi^+ \pi^-$, $K_{\rm S}^0 K^+ K^-$, $K_{\rm S}^0\pi^+ \pi^- \pi^0$ or $\pi^+ \pi^- \pi^+ \pi^-$. The framework is implemented in a model-independent method via a binned Dalitz plot analysis of the $D$ final state.


\section{Charm inputs}

The global average of $\gamma$ is driven mostly by inputs from ADS and GGSZ measurements. The ADS method needs inputs from $D$ decays: $r_{D}$ and $\delta_{D}$, the ratio of suppressed and favoured $D$ decay amplitudes and $D$ strong-phase respectively. For the ADS analysis of multibody $D$ decays, the coherence factor $\kappa$ is needed as input. It modulates the interference terms in the asymmetry parameters and varies between zero and one depending on whether there are many overlapping resonances contributing or a few isolated resonances. $D$ decay inputs $c_i$ and $s_i$, the amplitude weighted average of the cosine and sine of the strong phase difference between $D^0$ and $\bar{D^0}$ in different regions of phase space, are needed for a GGSZ style $\gamma$ extraction. New multibody $D$ modes can be explored in GLW framework, if $\mathit{CP}$  content $F_+$ is known. The interference terms in the asymmetry parameters will be modulated by $(2F_+-1)$.

 As the results are statistically limited, measuring these charm inputs from $B$ data leads to further loss in precision. So it is essential to measure them at a charm factory like CLEO-c where quantum-correlated $D\bar{D}$ mesons are produced in $e^{+}e^{-}$ collisions at an energy corresponding to $\psi(3770)$ resonance. The current CLEO-c inputs contributes 2$^\circ$ to the $\gamma$ uncertainty \cite{CLEO-unc}. In future, when $B$ statistics is expected to increase, the inputs from BES III experiment will be imperative.
 
\subsection{Quantum correlated $D$ mesons at CLEO-c} 

The $D$ meson pairs are produced coherently in a $C = -1$ state from $\psi (3770)$ decay. Thus the wave function becomes antisymmetric. The decay rate depends on the $\mathit{CP}$  eigenvalue of each $D$ final state. If they have opposite $\mathit{CP}$ , then there will be two-fold enhancement in the yield and the yield will be zero when both are of same $\mathit{CP}$ . It changes with them being quasi-$\mathit{CP}$  states or self-conjugate states. 

CLEO-c detector has a good 4$\pi$ solid angle coverage and hence full reconstruction of a $D\bar{D}$ event is possible. High efficiencies for track and photon reconstruction are also a feature of CLEO-c.

\section{Results}

A number of $D$ final states have been studied using a data sample corresponding to an integrated luminosity of $0.8~\mathrm{fb}^{-1}$  collected by CLEO-c. The $\mathit{CP}$  content, $c_i$ and $s_i$ values and coherence factor $\kappa$ have been measured for various multibody $D$ decay modes.

\subsection{$D \to K_{\rm S}^0\pi^+ \pi^-$}

\begin{figure}[ht!]
\begin{tabular}{c c}
\includegraphics[height=5cm, width=5cm]{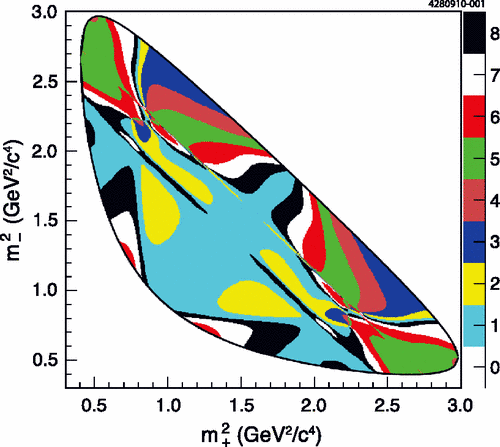} &
\includegraphics[height = 5cm,width= 6cm]{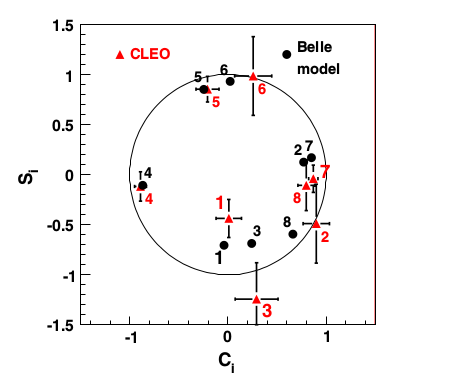}\\[0.5ex]
\end{tabular}
\caption{Binned Dalitz plot for $D \to K_{\rm S}^0\pi^+ \pi^-$ (left) and measured $c_i$ and $s_i$ values (right).} \label{Fig:Kspipi}
\end{figure}

This is the golden mode to measure $\gamma$ via the GGSZ formalism, especially at the $B$ factories. The $D$ Dalitz plot is binned and $c_i$ and $s_i$ values have been extracted in each of them as shown in Fig.~\ref{Fig:Kspipi} \cite{Kspipi}. Optimal binning, in terms of sensitivity to $\gamma$, is done with guidance from an amplitude model \cite{BaBar}. Eight symmetric bins have been constructed with respect to $m_{+}^2$ = $m_{-}^2$ line, where $m_{\pm}$ corresponds to invariant mass of $K_{\rm S}^0 \pi^{\pm}$.

These inputs have been used by both Belle and LHCb for $\gamma$ measurements \cite{Kspipi:Belle,Kspipi:LHCb}. In the model independent approach, the uncertainty due to modelling is replaced by the statistical uncertainty from CLEO-c sample.

\subsection{$D \to K_{\rm S}^0 \pi^+ \pi^- \pi^0$}

\begin{figure}[ht!]
\begin{tabular}{c c}
\includegraphics[height=4cm, width=5cm]{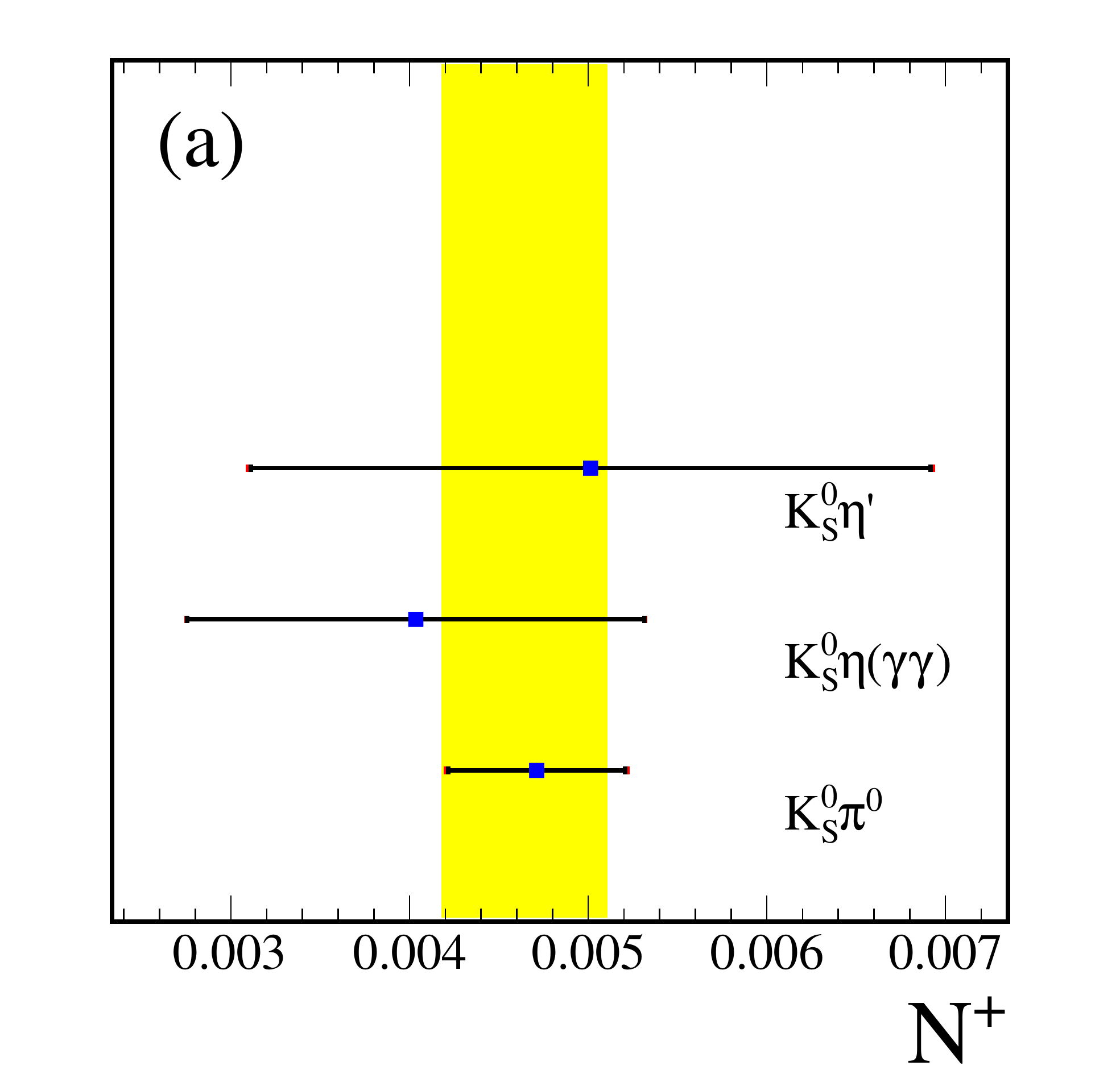} &
\includegraphics[height = 4cm,width= 5cm]{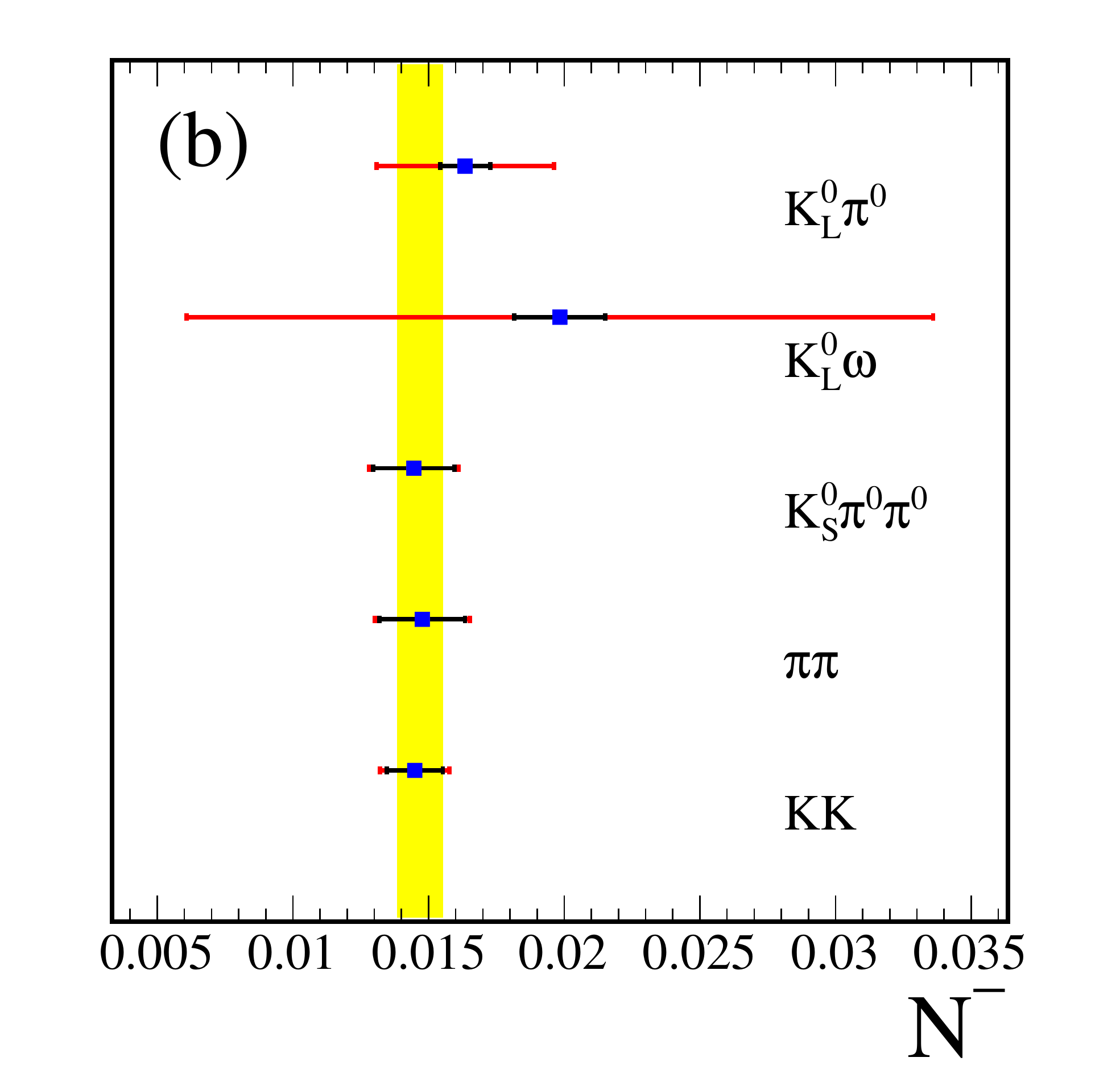}\\[0.5ex]
\end{tabular}
\caption{$N^+$ (left) and $N^-$ (right) for $D \to K_{\rm S}^0 \pi^+ \pi^- \pi^0$ with $\mathit{CP}$ -odd and even tags, respectively. Yellow region shows the average value.} \label{Fig:Kspipipi0N+}
\end{figure}

The decay $D \to K_{\rm S}^0 \pi^+ \pi^- \pi^0$ has a relatively large branching fraction of 5.2\% \cite{PDG}. This has been analysed against several tag (other $D$) modes that are $\mathit{CP}$  eigenstates, self-conjugate states $etc$. The yields with $\mathit{CP}$ -odd and $\mathit{CP}$ -even tags have been measured as $N^+$ and $N^-$ as given in Fig.~\ref{Fig:Kspipipi0N+}.

\begin{figure}[ht!]
\begin{tabular}{c c}
\includegraphics[height=4cm, width=5cm]{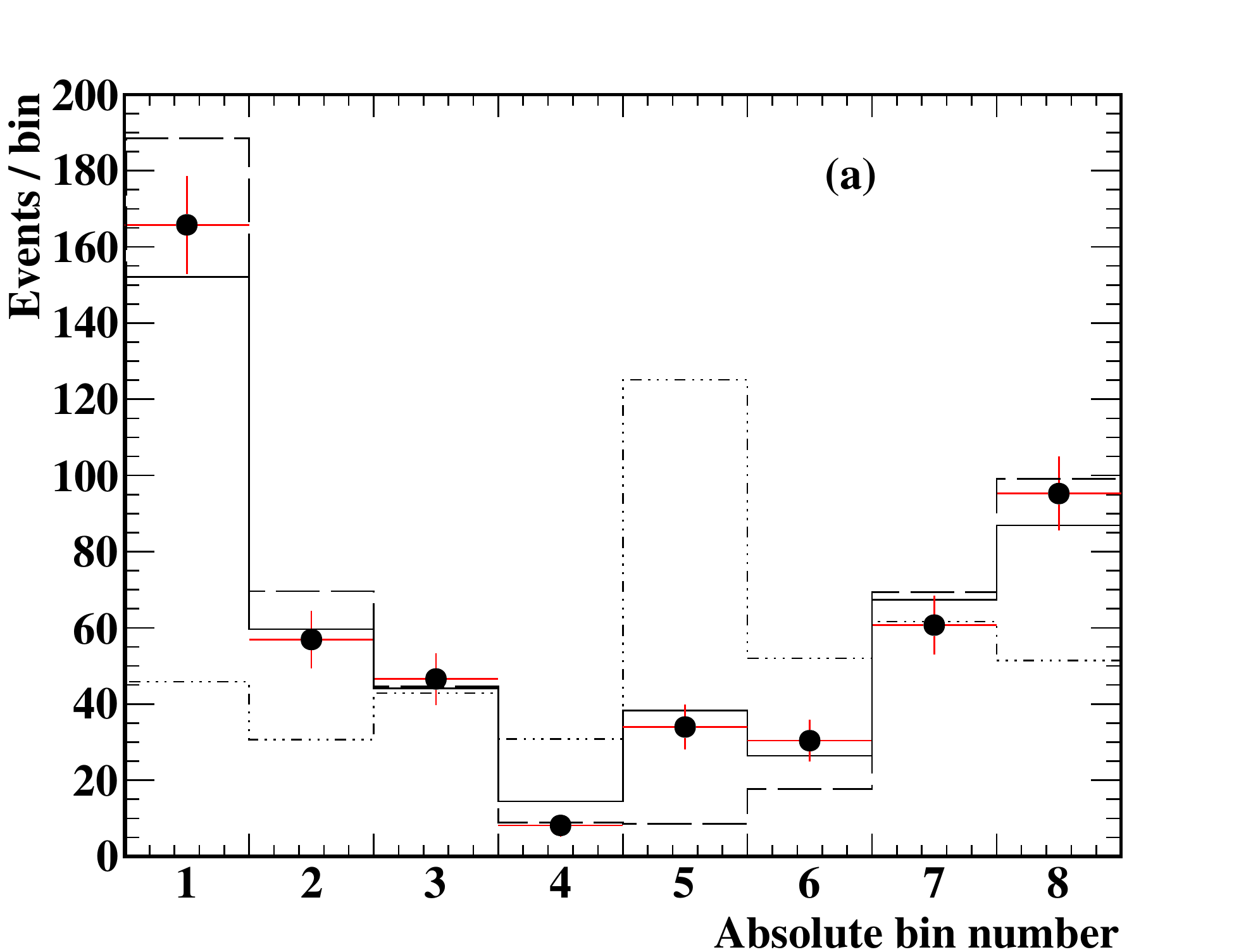} &
\includegraphics[height = 4cm,width= 5cm]{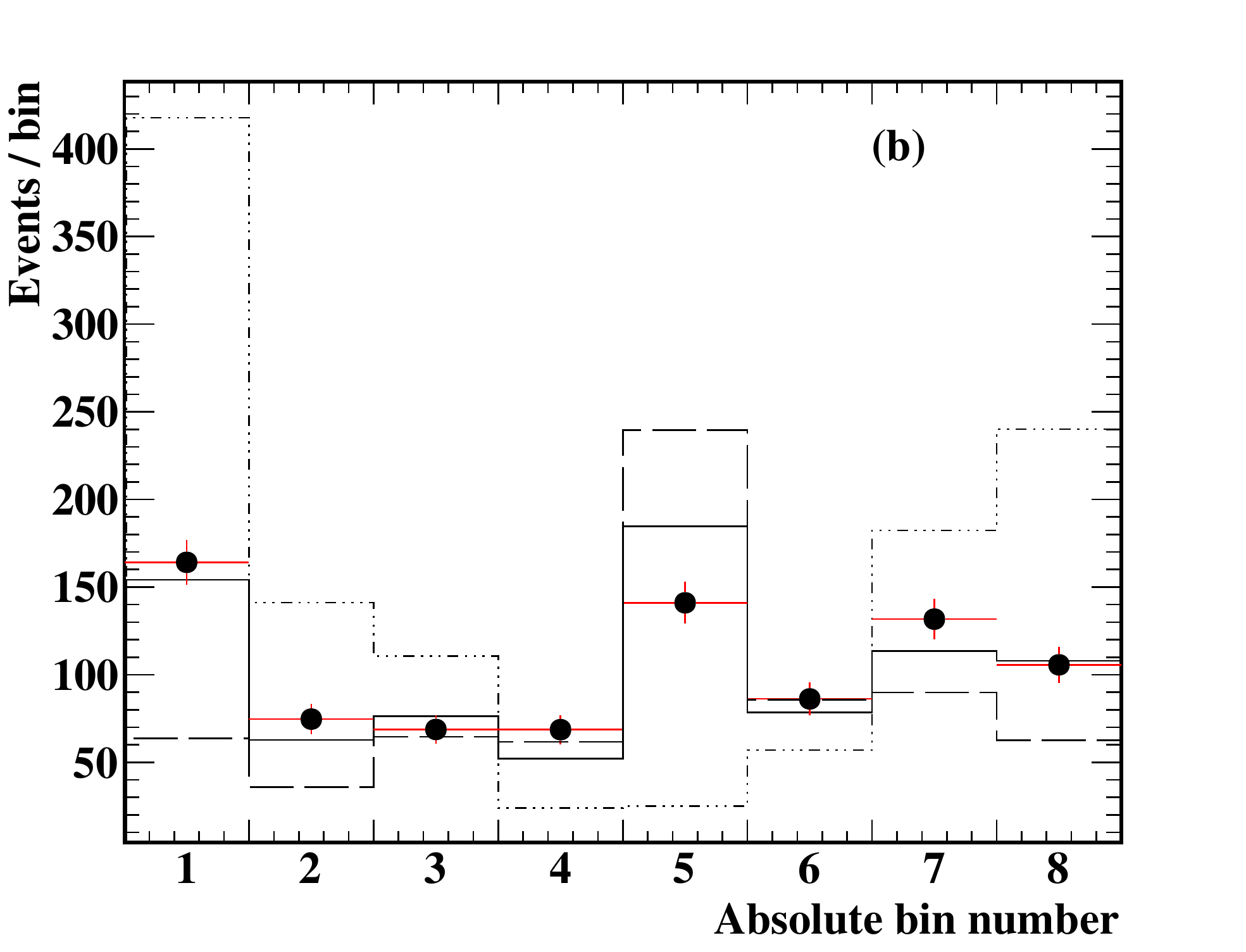}\\[0.5ex]
\end{tabular}
\caption{Measured and expected yields for $D \to K_{\rm S}^0 \pi^+ \pi^- \pi^0$ in $K_{\rm S}^0 \pi^+ \pi^-$ (left) and $K_{\rm L}^0 \pi^+ \pi^-$ (right) tag bins.} \label{Fig:kspipibins}
\end{figure}

The $\mathit{CP}$  content is defined as 
\begin{equation}
F_{+} = \frac{N^+}{N^+ + N^-}
\end{equation}
for $\mathit{CP}$  eigenstate tags. $K_{\rm S,L}^0\pi^+ \pi^-$ modes are used as self-conjugate tags. The yield measured in bins of tag Dalitz space is proportional to $1 - (F_+^{\rm sig} - 1)(F_{+}^{\rm tag} - 1)$. The measured and expected yields in $K_{\rm S,L}\pi^+ \pi^-$ bins are given in Fig.~\ref{Fig:kspipibins}.
From both the methods, the average $F_+ = 0.238 \pm 0.018$ \cite{Kspipipi0}, where the uncertainty includes both statistical and systematic contributions. This suggests that $K_{\rm S}^0 \pi^+ \pi^- \pi^0$ is significantly $\mathit{CP}$ -odd.

\begin{figure}[ht!]
\centering
\includegraphics[width=4cm, height=4cm]{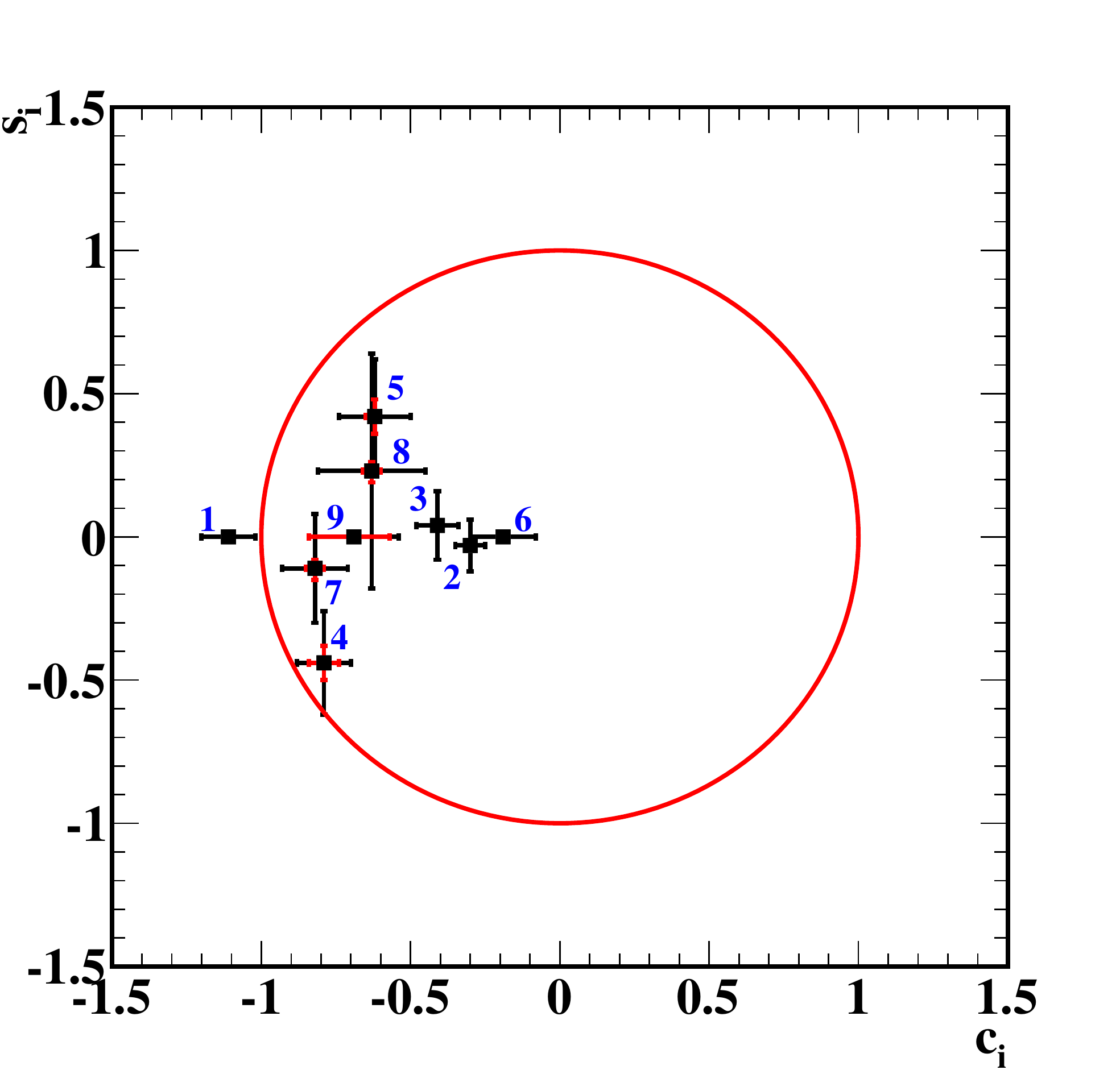}
\caption{$c_{i}$ and $s_{i}$ results in various bins for $K_{\rm S}^0 \pi^+ \pi^- \pi^0$.}\label{Fig:cisi}
\end{figure}

The multibody decay proceeds via some interesting resonance substructures. They are $\mathit{CP}$  eigenstates like $K_{\rm S}^0 \omega$ (GLW like), Cabibbo favoured states like $K^{*-}\pi^+ \pi^0$ (ADS like) $etc$. The strong-phase information can be extracted from the five dimensional phase space. In the absence of an amplitude model, the phase space is binned around these resonances. The results are given in table~\ref{Table:cisi} and Fig.~\ref{Fig:cisi} \cite{Kspipipi0}.

\begin{table} [ht!] 
\centering
\begin{tabular} {|c|c | c| c|} 
\hline 
Bin & resonance & $c_{i}$ & $s_{i}$\\[0.5ex]
\hline
\hline
1 &  $\omega$ &       $-1.11\pm0.09_{-0.01}^{+0.02}$ & 0.00\\[0.5ex]
2 &  $K^{*-}\rho^{+}$ &   $-0.30\pm 0.05 \pm 0.01$ & $-0.03\pm 0.09_{-0.02}^{+0.01}$\\[0.5ex]
3 &  $K^{*+}\rho^{-}$&       $ -0.41\pm0.07_{-0.01}^{+0.02}$ & $0.04\pm0.12_{-0.02}^{+0.01}$\\[0.5ex]
4 &  $K^{*-}$ &       $-0.79\pm0.09\pm 0.05$ &$-0.44\pm0.18\pm0.06$\\[0.5ex]
5 &  $K^{*+}$ &         $-0.62\pm0.12_{-0.02}^{+0.03}$ & $0.42\pm0.20\pm0.06$ \\[0.5ex]
6 &  $K^{*0}$ &        $-0.19\pm0.11\pm 0.02$ & 0.00\\[0.5ex]
7 &   $\rho^{+}$ &        $-0.82\pm0.11\pm 0.03$ & $-0.11\pm0.19_{-0.03}^{+0.04}$\\[0.5ex]
8 &  $\rho^{-}$ &         $-0.63\pm0.18\pm 0.03$& $0.23\pm0.41_{-0.03}^{+0.04}$ \\[0.5ex]
9 &    remainder &       $-0.69\pm0.15_{-0.12}^{+0.15}$ & 0.00\\[0.5ex]
\hline
\end{tabular}
\caption{$c_i$ and $s_i$ results in various bins for $K_{\rm S}^0 \pi^+ \pi^- \pi^0$.  Bins 1, 6 and 9 are $\mathit{CP}$  self-conjugate, which implies $s_{i}$ = 0.}\label{Table:cisi}
\end{table}

\begin{figure}[ht!]
\centering
\includegraphics[width=4cm, height=4cm]{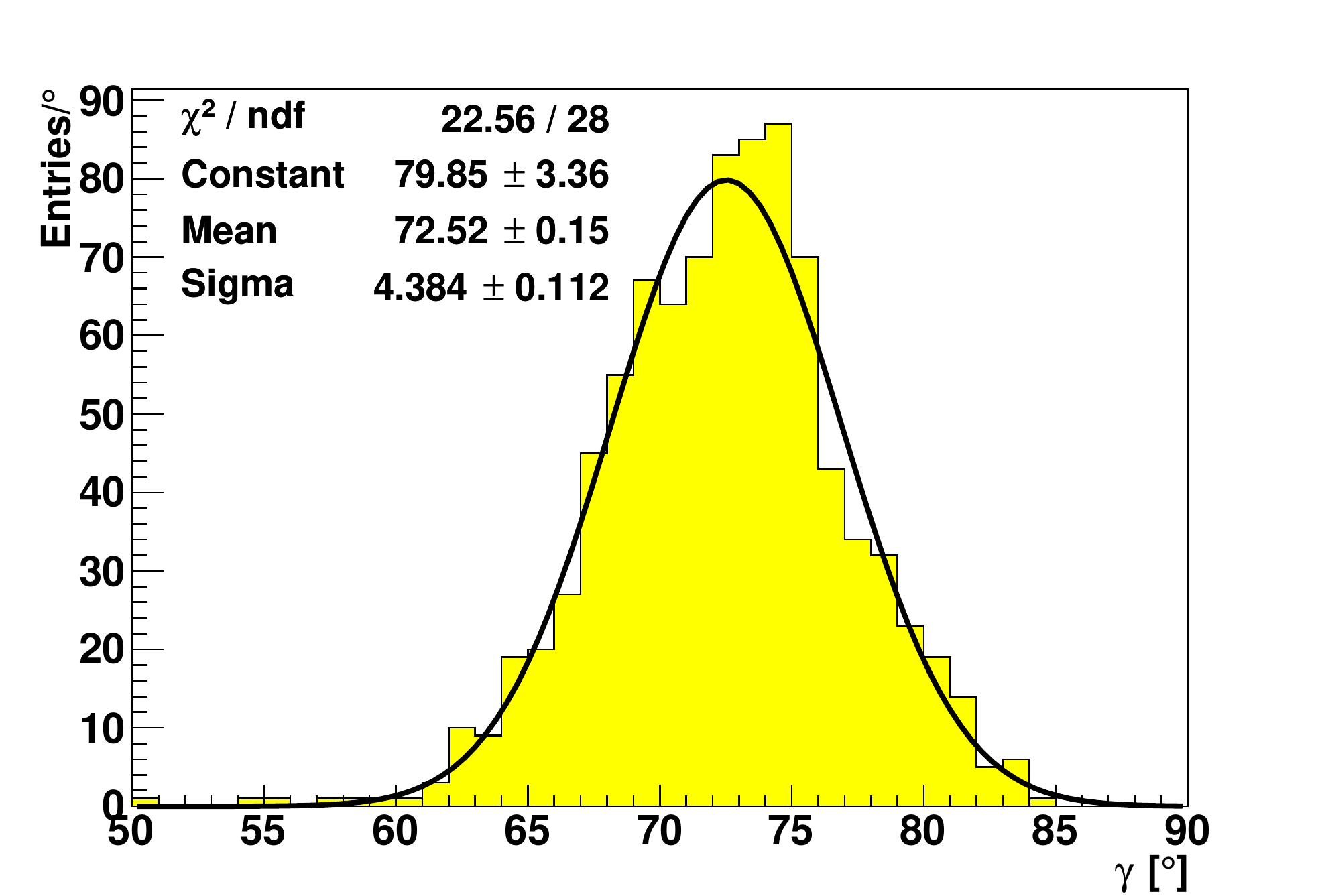}
\caption{$\gamma$ sensitivity with 50 ab$^{-1}$ dataset at Belle II.}\label{Fig:sensi}
\end{figure}

The $\gamma$ sensitivity with these results have been estimated for the expected full data set of 50 ab$^{-1}$ at Belle II. With $B^{\pm}\to D(K^{0}_{\rm S}\pi^{+}\pi^{-}\pi^{0})K^{\pm}$ decays, it is possible to reach $\sigma_{\gamma} =4.4^{\circ}$ (see Fig.~\ref{Fig:sensi}) \cite{Kspipipi0}.
Here, it is assumed that branching fraction $\times$ efficiency is similar to that of $K_{\rm S}^0 \pi^+ \pi^-$. Improvements are possible with optimal binning using the knowledge of an amplitude model or finer binning from a large statistics sample at BES III.

\subsection{$D \to \pi^+ \pi^- \pi^+ \pi^-$}

\begin{figure}[ht!]
\centering
\includegraphics[width=10cm, height=3cm]{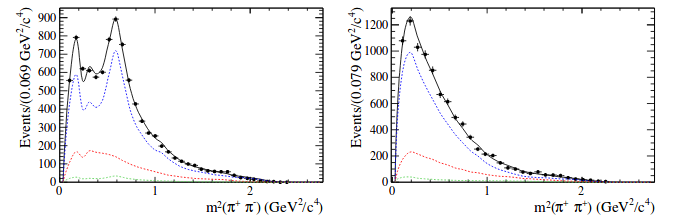}
\caption{Invariant mass projections for $\pi \pi$ combinations.}\label{Fig:4piamp}
\end{figure}

The all charged final state of $D \to \pi^+ \pi^- \pi^+ \pi^-$ makes the detection easier at LHCb. The phase space is analysed to extract $c_i$ and $s_i$ results. Here, binning is done based on amplitude model \cite{4pi-ampli}. The prominent contributions are $a_1 (1260)^+$ and $\rho (770)^0$. The invariant mass projections are shown in Fig.~\ref{Fig:4piamp}.

\begin{figure}[ht!]
\centering
\includegraphics[width=5cm, height=4cm]{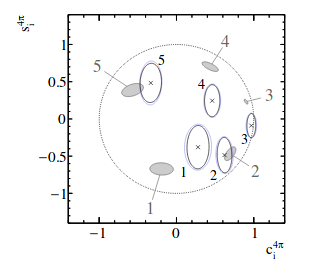}
\caption{$c_i$ and $s_i$ results for $D \to \pi^+ \pi^- \pi^+ \pi^-$. The grey regions shows the model predictions and black ellipses are measured values with statistical uncertainties.}\label{Fig:4picisi}
\end{figure}

\begin{figure}[ht!]
\centering
\includegraphics[width=6cm, height=4cm]{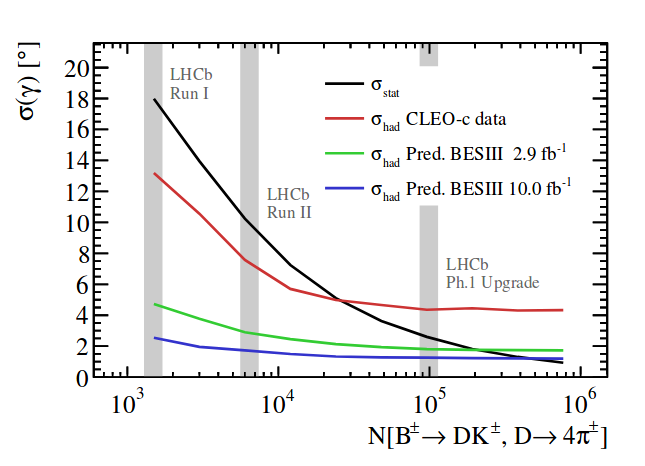}
\caption{Sensitivity predictions with $c_i$ and $s_i$ results for $D \to \pi^+ \pi^- \pi^+ \pi^-$.}\label{Fig:4pi_sensi}
\end{figure}

The five dimensional phase space is binned with the variables $m_{+}$, $m_{-}$, $\cos \theta_+$, $\cos \theta_-$ and $\phi$, where $m_{+} (m_-)$ is the invariant mass of $\pi^+ \pi^+ (\pi^- \pi^-)$ pair, $\theta_+ (\theta_-)$ is the helicity angle of $\pi^+ \pi^+ (\pi^- \pi^-)$ pair and $\phi$ is the angle between $\pi^+ \pi^+$ and $\pi^- \pi^-$ decay planes. The results with optimal binning are given in Fig.~\ref{Fig:4picisi} \cite{4pi-cisi}.
These results will contribute to 5$^{\circ}$ uncertainty on $\gamma$ with 50 fb$^{-1}$ data after LHCb phase I upgrade along with 2$^{\circ}$ from $B$ sample statistics. The sensitivity for different scenarios are presented in Fig.~\ref{Fig:4pi_sensi} \cite{4pi-cisi}.

\begin{figure}[ht!]
\centering
\includegraphics[width=10cm, height=4cm]{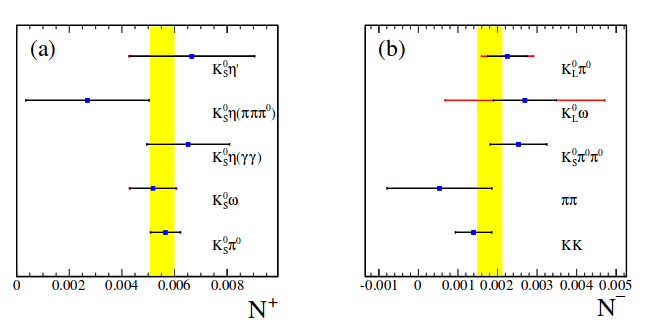}
\caption{$N^+$ (left) and $N^-$ (right) for $D \to \pi^+ \pi^- \pi^+ \pi^-$ with $\mathit{CP}$ -odd and even tags, respectively. Yellow region shows the average value.}\label{Fig:4pi_N}
\end{figure}

The $\mathit{CP}$  content $F_+$ has also been measured for this mode with $\mathit{CP}$  eigenstates and $K_{\rm S,L}^0 \pi^+ \pi^-$ modes as tags. The $N^+$ and $N^-$ results are shown in Fig.~\ref{Fig:4pi_N}.
The average $F_+$ result obtained is $0.737 \pm 0.028$ \cite{4pi}. Consistent results have been obtained using amplitude model as well as $c_i$ and $s_i$ results \cite{4pi-ampli,4pi-cisi}.

\subsection{$D \to K^- \pi^+ \pi^- \pi^+$}

\begin{figure}[!htb]
\centering
\includegraphics[width=5cm, height=4cm]{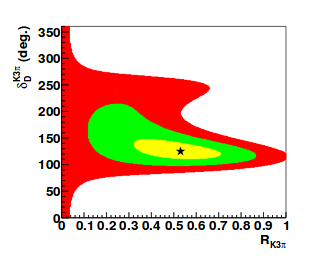}
\caption{Scans of $\Delta \chi^2$ for the fit to CLEO-c observables in $R_{K3\pi}, \delta_{D}^{K3\pi}$ parameter space.}\label{Fig:K3pi}
\end{figure}

The decay $D \to K^- \pi^+ \pi^- \pi^+$ can be analysed in ADS formalism to extract $\gamma$ if coherence factor $\kappa$ or $R_{K3\pi}$ is measured. It will then be treated like a two-body with single effective strong phase $\delta_{D}$. The results obtained with CLEO-c data are shown in Fig.~\ref{Fig:K3pi} \cite{k3pi}.

It has been also shown that $D - \bar{D}$ mixing results can be used as input for $\gamma$ measurements~\cite{k3pi_1}. So charm mixing results for this mode at LHCb \cite{k3pi_mix} are combined with these to obtain more precise values. There exists an amplitude model for $D \to K^- \pi^+ \pi^- \pi^+$ decay \cite{k3pi_amp} and this will allow for a more precise measurement of $\gamma$ by binning the five dimensional phase space. There is also scope for improvement to have a much more precise measurement of the coherence factors in those bins with more data at BES III.

\subsection{$D \to h^+ h^- \pi^0~~ (h=\pi, K)$ }

The $D \to \pi^+ \pi^- \pi^0$ Dalitz plot is symmetric and it suggests that it is isospin I = 0 state. Further, $G$-parity suggests that it is an almost pure $\mathit{CP}$ -even eigenstate. So this could potentially be used for $\gamma$ extraction in GLW method. It is analysed in CLEO-c data with $\mathit{CP}$  eigenstates and $K_{\rm S,L}^0 \pi^+ \pi^-$ modes as tags. The measured value of $F_+$ is $0.973 \pm 0.017$ \cite{MNayak,4pi}, which confirms that it is an almost pure $\mathit{CP}$ -even eigenstate. Similar measurement has been done for $D \to K^+ K^- \pi^0$ resulting in $F_+ = 0.732 \pm 0.055$ \cite{MNayak,4pi}. 

The coherence factor has been measured for $D \to K^- \pi^+ \pi^0$ decay mode to be $R_{K\pi\pi^0} = 0.82 \pm 0.06$ \cite{k3pi}. Since the value is close to 1, the dilution in the ADS observables due to strong phase from $D$ decay multi-particle phase space is quite small.

\section{Conclusions}
Quantum correlated $D$ decays at CLEO-c have been exploited to measure charm inputs that are used to determine $\gamma$ from $B$ meson decays. Inputs for GGSZ framework, $c_i$ and $s_i$, have been measured for $K_{\rm S}\pi^{+}\pi^{-}$, $K_{\rm S}^{0}\pi^{+}\pi^{-}\pi^{0}$ and $\pi^{+}\pi^{-}\pi^{+}\pi^{-}$ final states. Also, the coherence factor is measured for $K^{-}\pi^{+}\pi^{+}\pi^{-}$, which can be used in an ADS formalism to extract $\gamma$. The $\mathit{CP}$  content of decays like $\pi^{+}\pi^{-}\pi^{0}$, $K^{+}K^{-}\pi^{0}$ and $\pi^{+}\pi^{-}\pi^{+}\pi^{-}$ allow for new additions in GLW method. The precision on $\gamma$ will reach $\mathcal{O}(1^{\circ})$ with LHCb upgrade and Belle II. Therefore, inputs from BES III are required to prevent these measurements being systematically limited by uncertainties in the strong-phase measurements of $D$ decays.

%
%

\end{document}